\documentclass[11pt,fleqn,twosides]{article}

\usepackage{latexsym}
\usepackage[tbtags]{amsmath}
\usepackage{amsopn,amsthm,amsfonts,amssymb}
\usepackage{fancyhdr}
\usepackage{latexsym}
\fancypagestyle{plain}{ \fancyhead{} \fancyfoot{}

}

\newcommand{\mc}[1]{\ensuremath{\mathcal{#1}}}

\newcommand{\beq}{\begin{equation}}
\newcommand{\eeq}{\end{equation}}
\newcommand{\beqa}{\begin{eqnarray}}
\newcommand{\eeqa}{\end{eqnarray}}

\newtheorem{theorem}{Theorem}

\newtheorem{corollary}[theorem]{Corollary}

\newtheorem{definition}[theorem]{Definition}

\theoremstyle{remark}

\newtheorem*{acknow}{Acknowledgments}

\let\n\noindent
\let\la\lambda
\def\ds{\displaystyle}

\let\la\lambda
\let\La\Lambda

\let\t\theta
\let\om\omega

\let\Om\Omega
\let\n\noindent

\begin{document}
\small

\title{Explicit formulas for the generalized Hermite polynomials in
superspace}

\author{Patrick Desrosiers\thanks{pdesrosi@phy.ulaval.ca} \cr
\emph{D\'epartement de Physique, de G\'enie Physique et
d'Optique},\cr Universit\'e Laval, \cr Qu\'ebec, Canada, G1K 7P4.
\and Luc Lapointe\thanks{lapointe@inst-mat.utalca.cl }\cr
\emph{Instituto de Matem\'atica y F\'{\i}sica},\cr Universidad de
Talca,\cr Casilla 747, Talca, Chile.
     \and
     Pierre
Mathieu\thanks{pmathieu@phy.ulaval.ca} \cr \emph{D\'epartement de
Physique, de G\'enie Physique et d'Optique},\cr Universit\'e
Laval, \cr Qu\'ebec, Canada, G1K 7P4. }

\date{September 2003}

\maketitle

\begin{abstract}

We provide explicit formulas for the  orthogonal eigenfunctions of
the supersymmetric extension of the rational
Calogero-Moser-Sutherland model  with harmonic confinement, i.e.,
the generalized Hermite (or Hi-Jack) polynomials in superspace.
The construction relies on the triangular action of the
Hamiltonian on the supermonomial basis. This translates into
determinantal expressions for the Hamiltonian's eigenfunctions.

\end{abstract}

\newpage


\vfill\break

\section{Introduction}

The supersymmetric extension of the Calogero-Moser-Sutherland
(CMS) model \cite{CMS} has been obtained some thirteen years ago,
first in its rational version with confinement \cite{Freedman},
and then a few years later in its  trigonometric form \cite{SS}.
It took more than ten years before a single example of an
eigenfunction (for an arbitrary number of variables) could be
displayed for either model \cite{ghosh}, in which case the results
relied on a special feature of the rational model, namely the
existence of a transformation to a set of decoupled supersymmetric
oscillators.

This low progress is to be ascribed to the lack of a number of
crucial tools: 1) a general characterization of the symmetry
properties of the eigenfunctions; 2) a proper labelling of the
eigenfunctions; 3) the knowledge of a simple basis in terms of
which the eigenfunctions could be expanded. It is only quite
recently that the relevant tools have been designed \cite{DLM1}.
In that work, a systematic procedure for constructing the
eigenfunctions of the supersymmetric trigonometric (written stCMS)
model was presented and a number of solutions were displayed. In a
second step, these results were substantially improved by the
explicit construction of the eigenfunctions \cite{DLM2}. Finally,
the way to linearly combine these eigenfunctions so as to form a
set of orthogonal eigenfunctions was given in \cite{DLM3}. The
resulting eigenfunctions were dubbed Jack superpolynomials or
equivalently, Jack superpolynomials in superspace.  The superspace
referred to is the euclidian space of $N$ commuting variables (the
usual `positions' in the CMS model) augmented by the $N$
anticommuting variables generated by the
supersymmetrization.\footnote{These works are surveyed in
\cite{DLM4}.}

Having completely solved the trigonometric case, we then turned our
attention to
the rational model.
  A first construction of its eigenfunctions, called the generalized
Hermite  superpolynomials, was presented in \cite{DLM5}.  It is
the supersymmetric extension of Lassalle's result \cite{Las} in
that it relates the generalized Hermite  superpolynomial
$J^\om_\La$ to the Jack superpolynomial $J_\La$ (where $\La$ is a
superpartition and $\omega$ stands for the confining coefficient
-- see section 2) through the relation
\begin{equation}
      J_\La^\om= e^{-\Delta/4\om} J_\La\, .
\end{equation}
Here $\Delta$ is a certain differential operator of degree minus $2$ in
the commuting variables.
      Although very elegant, this construction does not appear, a
priori, to be a
practical way of constructing the generalized Hermite
superpolynomials. For instance, simply generating  the (ordinary)
polynomial $J_{(3,2,1)}^\om$ requires the calculation of $\Delta^3
J_{(3,2,1)}$.  Moreover, this kind
of computation has the disadvantage of depending on the number $N$
of variables in each of the two sets (commuting and anticommuting). In
addition, it does not directly express  the $J_\La^\om$'s in a
given basis of symmetric functions, e.g., the supermonomials
$m_\La$ or Jack polynomials $J_\La$.

In this article, we present a completely different construction of these
eigenfunctions.  We essentially apply the scheme used in \cite{DLM2} and
thereby obtain  explicit expressions for the $J_\La^\om$'s in terms of
supermonomials. This method is exposed in full generality in
appendix A (which can be read independently of this article).  The
underlying  idea is that there is a road to the construction
of the eigenfunctions whose central step is the solution of a very simple
two-body problem.  The motivation for this work was to present
a practical construction of the srCMS model (i.e., free of the
previously mentioned drawbacks of the method exposed in \cite{DLM5})
and at the same time to advertise the power of this general scheme.

The article is organized as follows. We first review briefly some
(super)material needed for our construction. In section 3, we
evaluate the action of the Hamiltonian on the supermonomial basis
$m_\La$. The calculation is broken down into two parts: we first
evaluate the action of the Hamiltonian in the one- or two-particle
sector (the former case taking care of the diagonal part of the
Hamiltonian) and then extend the result to a general number of
particles by adding the right symmetry factors. Breaking down the
computation in this manner is obviously the only way of
unravelling a clear pattern for the action of the Hamiltonian on
the supermonomials and, in the end, for the triangular
decomposition of the generalized Hermite superpolynomials in terms
of Jack superpolynomials.
Each coefficient $c_{\La,\Om}$ can thus be
written as a sum of terms, each dressed by their respective
symmetry factor.

Once the action of srCMS Hamiltonian on the supermonomials is
completely fixed, all the data necessary for computing the
generalized Hermite superpolynomials are known.   In Section 4, we
end up with a tri- determinantal expression for generalized
Hermite superpolynomials in terms of the supermonomials. We stress
that even in the zero-fermion sector, the resulting expressions
are new (albeit implicit in \cite{Las}).\footnote{For more details
concerning the generalized Hermite polynomials without fermions,
see also \cite{vandiejen} and \cite{BF}.}

The appendix B completes the article by providing the
determinantal formulas related to the Jack superpolynomials. From
     a computational view point, this essentially summarizes the results
obtained in Refs
     \cite{DLM2,DLM3}.


\section{Definitions}

The supersymmetric gauge transformed Hamiltonian of the srCMS
model \cite{Freedman} reads:
\begin{equation} \label{shjack}
\mc{H}= 2\om\sum_{i=1}^N (x_i
\partial_i+\theta_i\partial_{\theta_i})- \sum_{i=1}^N
\partial_i^2-2\beta \sum_{1\leq i<j\leq N}\frac{ 1} {x_{ij}}(
\partial_i-\partial_j)+2\beta\sum_{1\leq i<j\leq
N}\frac{1}{x_{ij}^2}(\theta_{ij}\theta^\dagger_{ij})
\, ,
\end{equation}
where
     \beq
\partial_i= \partial_{x_i},\qquad x_{ij}=x_i-x_j,\qquad\theta_{ij} =
\theta_{i} -
\theta_{j}\qquad\mbox{and}\qquad \theta_{ij}^{\dagger} =
\partial_{\theta_{i}} - \partial_{\theta_{j}}.\eeq
     The
$\theta_{i}$'s, with  $ i=1,\cdots,N $, are anticommuting
(Grassmannian or fermionic) variables.  The commuting (bosonic)
variables $x_j$ as well as the parameters $\beta$ and $\om$ belong
to the real field. It is easy to verify that the term
$1-\theta_{ij}\theta^{\dagger}_{ij}$ is a fermionic exchange
operator \cite{SS}
\begin{equation}
\kappa_{ij}\equiv 1-\theta_{ij}\theta^{\dagger}_{ij}=
1-(\theta_{i}-\theta_j)(\partial_{\theta_i}-\partial_{\theta_j})
\, ,
\end{equation}
whose action on a function
        $ f(\theta_i,\theta_j) $ is
\begin{equation}
\kappa_{ij}\, f(\theta_i,\theta_j)= f(\theta_j,\theta_i)\,
\kappa_{ij}\, .
\end{equation}
The Hamiltonian $\mc{H}$ is Hermitian with respect to the
following `physical scalar product': \beq
\label{defscaprod}\langle
F(x,\theta),G(x,\theta)\rangle_{\beta,\omega}=\prod_{j}\left(\int_{-\infty}^\infty
dx_j\int d\theta_j\theta_j\right) \prod_{k\leq
l}(x_{kl})^{2\beta}e^{-\omega\| x \|^2}
F(x,\theta)^*G(x,\theta),\eeq where $F$ and $G$ are arbitrary
functions and where $\| x \|^2=\sum_i x_i^2$.  The complex
conjugation $*$ is defined such that\beq\label{defcomplex}
(\theta_{i_1}\cdots\theta_{i_m})^*\theta_{i_1}\cdots\theta_{i_m}=1\qquad\mbox{and}\qquad
x_j^*=x_j.\eeq In other words, $\theta_j^*$ behaves  as
$\theta_j^\dagger=\partial_{\theta_j}$.  The integration over the
Grassmannian variables is the standard Berezin integration, i.e.,
\beq \int d\theta=0,\qquad\int d\theta \,\theta=1.\eeq

The Hamiltonian  $\mc{H}$ preserves the space, $P^{S_N}$,
of symmetric
superpolynomials invariant under the simultaneous
action of $ \kappa_{ij} $  and  $ K_{ij} $, where  $ K_{ij} $ is
the exchange operator acting on the  $ x_i $ variables:
\begin{equation}
               K_{ij}f(x_i, x_j)=f(x_j, x_i)K_{ij} \, .
\end{equation} A polynomial $f$ thus belongs to $P^{S_N}$ if it is
invariant under the action of the product
      \beq \mc{K}_{ij}= \kappa_{ij}K_{ij} \, , \eeq
that is, if $\mc{K}_{ij} f=f$ for all $i,j$.

The appropriate labelling for symmetric superpolynomials is
provided by  superpartitions \cite{DLM1}. We recall that a
superpartition $\Lambda$ in the $ m $-fermion sector is a sequence
of non-negative integers composed of two standard partitions
separated by a semicolon: \beq \Lambda=(\Lambda^a
;\Lambda^s)=(\Lambda_1,\ldots,\Lambda_m;\Lambda_{m+1},\ldots,\Lambda_{N})\,
,\eeq
     where $ \Lambda_i>\Lambda_{i+1}\geq 0$ for $ i=1, \ldots
m-1$ and $\Lambda_j \ge \Lambda_{j+1}\geq 0 $ for
$j=m+1,\dots,N-1$. In the zero-fermion sector, the semicolon is
usually omitted and
        $ \Lambda $  reduces then to  $ \Lambda^s $. We denote the degree of a
superpartition and its fermionic number respectively by:
\begin{equation}
|\Lambda|=\sum_{i=1}^{N}\Lambda_i,\qquad\mbox{and}\qquad
\overline{\underline{\Lambda}} = m\, .
\end{equation}
The partition rearrangement in non-increasing order of the entries
of $\Lambda$ is denoted $\Lambda^*$.

This allows to define dominance (partial) ordering on
superpartitions.  We  recall the usual dominance ordering on two
partitions $\lambda$ and $\mu$ of the same degree: $\lambda \leq
\mu$ iff
${\lambda}_1+\ldots+{\lambda}_k\leq{\mu}_1+\ldots+{\mu}_k$ for all
$k$. The dominance ordering on superpartitions is
similar:\beq\label{dominanceorder} \Omega\leq\Lambda\quad\mbox{if
either}\quad\Omega^*<\Lambda^*\quad\mbox{or}\quad\Omega^*=\Lambda^*\quad\mbox{
and}\quad{\Omega}_1+\ldots+{\Omega}_k\leq{\Lambda}_1+\ldots{\Lambda}_k,\quad
\forall\,\, k\, .\eeq

We now furnish  two fundamental bases in the space $P^{S_N}$.  The
simplest one is the monomial basis, denoted $\{m_\La\}_\La$,
with $m_\La\equiv m_\La(x,\theta)$
a superanalogue of a monomial symmetric function
defined as follows \cite{DLM1}:
\begin{equation}
m_{\Lambda}={\sum_{\sigma\in S_{N}}}' \theta^{\sigma(1, \ldots,
m)}x^{\sigma(\Lambda)} \, ,
\end{equation} where the prime indicates that the  summation is
restricted to distinct terms, and where
\begin{equation}
x^{\sigma(\Lambda)}=x_1^{\Lambda_{\sigma(1)}} \cdots
x_m^{\Lambda_{\sigma(m)}} x_{m+1}^{\Lambda_{\sigma(m+1)}} \cdots
x_{N}^{\Lambda_{\sigma(N)}}\qquad\mbox{and}\qquad\theta^{\sigma(1,
\ldots, m)} = \theta_{\sigma(1)} \cdots \theta_{\sigma(m)} \, .
\end{equation}
Equivalently, we can define the supermonomials as
\begin{equation} \label{monod}
       m_{\Lambda}= \frac{1}{f_{\Lambda}} \sum_{\sigma \in S_N} \mc{K}_{\sigma}
       \left( \theta_1 \cdots \theta_m x^{\Lambda}\right) \, ,\qquad \quad
f_{\Lambda} =f_{\Lambda^s}= n_{\Lambda^s}(0)!\, n_{\Lambda^s}(1)!
\,
       n_{\Lambda^s}(2) ! \cdots \, ,
\end{equation}
where  $n_{\Lambda^s}(i)$ indicates  the number of $i$'s in
$\Lambda^s$,
       the symmetric part of $\Lambda=(\Lambda^a;\Lambda^s)$, and where
$\mc{K}_{\sigma}$ stands for $\mc{K}_{i_1,i_1+1} \cdots
\mc{K}_{i_n,i_n+1}$ when the element $\sigma$ of the symmetric
group $S_N$ is written  in terms of elementary transpositions,
i.e., $\sigma =
       \sigma_{i_1} \cdots
       \sigma_{i_n}$.

The second basis, denoted $\{J_\La\}_\La$,  is that of the
     superanalogues of the Jack polynomials \cite{DLM3}. These
superpolynomials are orthogonal and triangular eigenfunctions of
the gauged stCMS Hamiltonian $\mc{H}_2$.  More precisely,  the
Jack polynomials in superspace $J_\Lambda\equiv
J_\La(x,\theta;1/\beta)$  are the unique functions in $P^{S_N}$
such that
     \beq\label{defjack1}
J_\Lambda=m_\Lambda+\sum_{\Omega<\Lambda}t_{\Lambda,\Omega}(\beta)m_\Omega
\qquad\mbox{and}\qquad\langle
J_\Lambda,J_\Omega\rangle_{\beta}\propto \delta_{\Lambda,\Omega}
\, , \eeq where the physical scalar product of the stCMS model is
defined by \beq \langle
A(x,\theta),B(x,\theta)\rangle_{\beta}=\prod_{j}\left(\frac{1}{2\pi
i} \oint \frac{ dx_j}{x_j}\int
d\theta_j\,\theta_j\right)\left[\prod_{k\neq
l}\left(1-\frac{x_k}{x_l}\right)^\beta
A(x,\theta)^*B(x,\theta)\right],\eeq The variable $x_j$ represents
the $j^{\mbox{\tiny th}}$ particle's position on the unit circle
in the  complex plane, and is thus such that $x_j^*=1/x_j$.

We finally give two distinct types of eigenfunctions of the srCMS
Hamiltonian $\mc{H}$: the non-homogenous supermonomials
$m_\La^\om$ and the generalized Hermite superpolynomials
$J_\La^\om$ \cite{DLM5}. They are `$\om$-deformations' of the
supermonomials and the Jack superpolynomials respectively, in the
sense that
\beq
\lim_{\om\rightarrow\infty}m_\La^\om=m_\La\qquad\mbox{and}\qquad\lim_{\om\rightarrow\infty}J_\La^\om=J_\La\,
.\eeq The set of superpolynomials $\{m_\La^\om\}_\La$, where
$m_\La^\om=m_\La(x,\theta;\beta,\om)$, is the unique basis of
$P^{S_N}$ satisfying \beq\label{defmomega}
m_\La^\om=m_\La+\sum_{\Om<_u\La}y_{\La\Om}(\beta,\om,N)\,m_\Om\qquad{\rm {and}}
\qquad\mc{H}
m_\Lambda^\om=2\om \, (|\La|+\overline{\underline{\La}}) \, m_\La^\om \,
,\eeq where the $u$-ordering is  such that \beq\label{uorder}
\Om\leq_u\La \qquad\mbox{if either}\qquad \Om=\La
\qquad\mbox{or}\qquad |\Om|=|\La|-2n \, ,\eeq for $n=1,2,3,\ldots$ Note
that we compare superpartitions belonging to the same fermionic
sector, i.e., such that
$\overline{\underline{\La}}=\overline{\underline{\Om}}$.
Similarly, the generalized Hermite polynomials
$J^{\om}_{\Lambda}\equiv J_\La(x,\theta;1/\beta,\om)$ form the
unique basis of $P^{S_N}$ satisfying
     \beq \label{defhermite1}
J_\La^\om=J_\La+\sum_{\Om<_u\La}w_{\La\Om}(\beta,\om,N)\,J_\Om\qquad
\mbox{and}\qquad\mc{H}
J_\Lambda^\om=2\om \, (|\La|+\overline{\underline{\La}}) \, J_\La^\om \,
.\eeq  However, in contradistinction with the $m_\La^\om$'s, the
$J_\La^\om$'s are also orthogonal. Actually,
$\{J^{\om}_{\Lambda}\}_\La$ is the only basis in $P^{S_N}$ having
the two following  properties \cite{DLM5}:
     \beq\label{defhermite2}
     J_\La^\om=J_\La+\sum_{\Om<_u\La}w_{\La
\Om}(\beta,\om,N)\,J_\Om\qquad\mbox{and}\qquad \langle
J_\Lambda^\om,J_\Om^\om\rangle_{\beta,\om} \propto
\delta_{\La,\Om} \,.\eeq
with $ \langle
\;,\; \rangle_{\beta,\om}$ defined in  (\ref{defscaprod}).

The explicit form of the eigenfunctions $m_\La^\om$ and
$J_\La^\om$ will be obtained in the next two sections.



\section{Action of the srCMS Hamiltonian on supermonomials}

We want to compute the coefficients $c_{\La\Om}$ in the
development \beq\label{tringenm} \mc{H}
m_\La=e_{\La}m_\La+\sum_{\Om <_v \La}c_{\La\Om}m_\Om\eeq in terms
of the ordering $\leq_v$ to be introduced in the next subsection.

     Proceeding as in \cite{DLM2}, we divide the calculation of
$\mc{H}\,m_\La$ into two parts.  We first treat the special cases
$N=1,2$.   The central role played by the
      one-particle and two-particle sectors is rooted in the fundamental
observation that the Hamiltonian is a sum of one and two-particle
interactions.   This computation will provide the core of the
coefficient $c_{\Lambda\Omega}(\beta,\om)$ appearing in
(\ref{tringenm}). The consideration of the case $N>2$ will only
dress them by symmetry factors.

To simplify the calculations, we introduce  the following
notation:
\begin{equation} \label{shjacka}
\mc{H}=2\om\mathcal{A}-(\mathcal{B}+2 \beta \mathcal{C}) ,
\end{equation}
where
\begin{equation}
\mathcal{A}=\sum_i \mathcal{A}_i=  \sum_i
(x_i\partial_i+\theta_i\partial_{\theta_i}),\qquad \mathcal{B}=
\sum_i \mathcal{B}_i= \sum_i \partial_i^2
\end{equation}
and
\begin{equation}
\mathcal{C}= \sum_{i<j}\mathcal{C}_{ij}=
\sum_{i<j}\left[\frac{1}{x_{ij}}
(\partial_i-\partial_j)-\frac{1}{x_{ij}^2} (1-\kappa_{ij})
\right]\, .\end{equation}

     Only the term $\mathcal{A}$ has a diagonal part.
Its action is easily computed for all $N$:
\begin{equation}
2\om \mathcal{A}\,
m_\Lambda=e_{\La}\,m_\La=2\om(|\La|+\overline{\underline{\La}})\,m_\La.
\end{equation}

\subsection{Action of the operator $\mathcal{B}$; the $U_i$
ladder operator}

The operator $\mathcal{B}$ acts trivially on one-particle
supermonomials $m_{(;r)}=m_{(r)}=x_1^r$ and $m_{(r;)}=\theta_1
x_1^r$:
\begin{equation} \label{acdeb}
\mathcal{B}_1 m_{[r]}=r(r-1)m_{[r-2]}\, ,
\end{equation}
where $[r]$ stands for $(;r)$ or $(r;)$.

We translate the action of $\mathcal{B}_i$ on the supermonomial
$m_\Lambda$ into that of a {\it ladder operator} $ U_i$ acting on
the superpartition $\Lambda$ as :
\begin{equation}
U_i \La = U_i (\La_1,\cdots, \La_i,\cdots,\La_N)=(\La_1,\cdots,
\La_i-2,\cdots,\La_N) \, .
\end{equation}
We will denote by ${\rm{sgn}}(\sigma^a_{U_i\Lambda})$ the sign of
the permutation $\sigma^a_{U_i\Lambda}$ needed to reorder the
entries of $U_i\Lambda$ (on the antisymmetric side) in order to have
a superpartition:  $\overline{U_i
\La}=\sigma^a_{U_i\Lambda}(U_i \La)$.  Notice that $\La_i$ can
either be part of $\La^a$ or $\La^s$, but a sign can only arise if
$\La_i\in\La^a$.

Since $|\overline{U_i\La}|=|\La|-2$, successive applications of
the $U_i$ operators on a given superpartition generate a
specialization of the $u$-ordering (see Eq. (\ref{uorder})): \beq
\mbox{if}\qquad\Om=\overline{U_{i_1}\cdots\overline{U_{i_n}\Lambda}},
\qquad\mbox{for some } i_1,\dots,i_n, \qquad\mbox{then} \qquad
\Om\leq_u\La\, .\eeq

In the general case of $ N\geq2$, we have
\begin{equation}\label{defp}
\mc{B}\, m_\Lambda=\sum_{\Omega;
\Omega=\overline{U_i\Lambda}}d_{\La \Om} m_\Omega ,
\end{equation}
where the sum is taken only over the different superpartitions
$\Om=\overline{U_i\La}$, and where
\begin{equation}\label{pLaOm}
d_{\La
\Om}=\begin{cases}\Lambda_i(\Lambda_i-1){\rm{sgn}}(\sigma_{U_i\Lambda}^a)\,
,& \mbox{if} \quad i\in\{1,\ldots,m\}\, \\
\Lambda_i(\Lambda_i-1) n_{\Omega^s}(\Lambda_i-2)\, ,&\mbox{if}
\quad i\in\{m+1,\ldots,N\}\, .
\end{cases}
\end{equation}
Here $n_{\Omega^s}(\Lambda_i-2)$ is a symmetry factor (recall that
$n_{\Omega^s}(a)$ gives the number of $a$'s in $\Omega^s$).  This
non-trivial coefficient is determined using calculations similar
to those presented in Section 4.2 of Ref. \cite{DLM2}.  It
corresponds to $f_\Omega\#(\Om,\La)/f_\La$, where $\#(\Om,\La)$
denotes the number of distinct ways we can choose $i$ such that
$\Om=\overline{U_i\La}$ and $f_\La$ is the monomial coefficient
defined in Eq. (\ref{monod}).  For instance, if $\La=(0;2,1)$ then
$\Om=\overline{U_2(0;2,1)}=(0;1)$ has a symmetry factor
$f_\Omega\#(\Om,\La)/f_\La=(N-2)!\cdot 1/(N-1)!=(N-2)$, so $\ds
d_{(0;2,1)\,(0;1)}=2(N-2)$.

\subsection{Action of the operator $\mathcal{C} $ in the
two-particle sector }

The operator $\mc{C}$ contains the two-body interaction terms of
the Hamiltonian. To compute its action in the two-particle sector,
we need to distinguish three cases, characterized by their
different fermionic sector.\\

\n {\bf Case I}\\

In this case, the fermionic number is two and the supermonomial to
be considered is $m_{(r,s;0)}$:
\begin{equation}\label{lekaaa}
        m_{(r,s;0)}=\frac{1}{f_{(r,s;0)}}(1+\mc{K}_{12})\,
\theta_1 \theta_2 x_1^r x_2^s = \t_1\t_2(x_1^rx_2^s-x_1^s x_2^{r})
\, ,
\end{equation}
        for $r>s$. For the action of $\mc{C}$, a direct computation yields
\begin{eqnarray}
\mc{C}_{12} m_{(r,s;0)} = &&\frac{\t_1\t_2}{x_{12}}  (x_1
x_2)^{s-1} \{ r(x_1^{r-s}x_2+x_1x_2^{r-s})-
s(x_1^{r-s+1}+x_2^{r-s+1})\cr &&\qquad
\qquad\qquad-2(x_1^{r-s}x_2+x_1^{r-s-1}x_2^2+\cdots
+x_1x_2^{r-s})\} \, .
\end{eqnarray}
To proceed further, we use the following identity:\footnote{Similar
identities are presented in \cite{DLM2}.  Identity \eqref{iden48}
can be proved along the same lines. The way such identities are
found is rather simple however. The idea is to add an appropriate
monomial to each monomial treated successively  in order
       to be able to extract a factor $x_{12}$. For instance, for the
monomial with the leading power of $x_1$, we have:
       $ -sx_1^{r-s+1}= -s[ x_1^{r-s}x_{12}+ x_1^{r-s}x_2]$.
       We then collect all other terms of the form  $ x_1^{r-s}x_2$ in
our original
expression,
       add them to the newly produced $-s x_1^{r-s}x_2$ and proceed
similarly to enforce the factorization of $x_{12}$. }
\begin{eqnarray} \label{iden48}
        &&\{
r(x_1^{r-s}x_2+x_1x_2^{r-s})- s(x_1^{r-s+1}+x_2^{r-s+1})
-2(x_1^{r-s}x_2+x_1^{r-s-1}x_2^2+\cdots +x_1x_2^{r-s})\}
        \cr
&&\qquad \quad = x_{12} \big\{ -s(x_1^{r-s}-x_2^{r-s})
+\sum_{\ell=1}^{\lfloor(r-s-1)/2\rfloor}
(r-s-2\ell)(x_1^{r-s-\ell}x_2^\ell -x_1^\ell
x_2^{r-s-\ell})\big\}\, .
\end{eqnarray}
We then get
\begin{equation}\label{lafiaa}
\mc{C} m_{(r,s;0)}= \mc{C}_{12} m_{(r,s;0)}=-s\,m_{(r-1,s-1;0)}+
\sum_{\ell=1}^{ \lfloor (r-s-1)/2\rfloor }
(r-s-2\ell)\,m_{(r-1-\ell,s-1+\ell;0)}\, .\\
\end{equation}

\n{\bf Case II}\\

In this case, we work in the one-fermion sector with $m_{(r;s)}$:
\begin{equation}\label{lekaa}
m_{(r;s)}= \frac{1}{f_{(r;s)}}(1+\mc{K}_{12})\, \theta_1 x_1^r
x_2^s = \t_1x_1^rx_2^s+\t_2x_1^sx_2^r\, ,
\end{equation}
(since $f_{(r;s)}=1$). Consider  the action of $\mc{C}$ on
$m_{(r;s)}$, supposing first that $r>s$. We get
\begin{eqnarray}
C_{12}m_{(r;s)}&=& \frac{1}{x_{12}}  (x_1x_2)^{s-1} [\t_1
(rx_1^{r-s}x_2 -sx_1^{r-s+1})- \t_2 (rx_1x_2^{r-s} -sx_2^{r-s+1})
\cr &&\phantom{\frac{1}{x_{12}}  (x_1x_2)^{s-1} [} -\t_{12}
(x_1^{r-s}x_2+x_1^{r-s-1}x_2^{2} +\cdots + x_1 x_2^{r-s})]\, .
\end{eqnarray}
We now focus on the $\t_1$ term:
\begin{equation}
\left. \mc{C}_{12} m_{(r;s)}\right|_{\t_1}=
\frac{\t_1}{x_{12}}(x_1x_2)^{s-1}\{
rx_1^{r-s}x_2-sx_1^{r-s+1}-(x_1^{r-s}x_2 +\cdots +
x_1x_2^{r-s})\}\, .
\end{equation}
A factor $x_{12}$ can again be factorized from the curly bracket,
which can be seen using the
       identity
\begin{equation}
\{ rx_1^{r-s}x_2-sx_1^{r-s+1}-(x_1^{r-s}x_2 +\cdots +
x_1x_2^{r-s})\} = x_{12}\{ -sx_1^{r-s}+ \sum_{\ell=1}^{r-s-1}
(r-s-\ell)x_1^{r-s-\ell}x_2^\ell\}\, .
\end{equation}
Reinserting the $\t_2$ terms (by symmetry), we  have
\begin{equation}\label{lafia}
\mc{C} m_{(r;s)}=  \mc{C}_{12}m_{(r;s)}=  -s\, m_{(r-1;s-1)}+
\sum_{\ell=1}^{r-s-1}(r-s-\ell) \, m_{(r-1-\ell;s-1+\ell)} \qquad
(r>s)\,.
\end{equation}
The derivation in the case $r<s$ is similar, with the role of the
parts $r$ and $s$ interchanged:
\begin{equation}
\mc{C} m_{(r;s)}=  \mc{C}_{12}m_{(r;s)}=  -r\, m_{(r-1;s-1)}+
\sum_{\ell=1}^{s-r-1}(s-r-\ell) \, m_{(r-1+\ell;s-1-\ell)} \qquad
(r<s)\,.
\end{equation}
Here, no reordering of the partition is required.  The $r=s$ case
is trivial: \beq \mc{C} m_{(r;r)}=  \mc{C}_{12}m_{(r;r)}=  -r\,
m_{(r-1;r-1)}\, .\eeq

\n{\bf Case III}\\

      This case corresponds to the zero-fermion sector. We set
$\Lambda=(;r,s)=(r,s)$, so that
\begin{equation}\label{leka}
m_{(r,s)}= \frac{1}{f_{(r,s)}}(1+\mc{K}_{12})\, x_1^r x_2^s =
\frac{1}{f_{(r,s)}} ( x_1^rx_2^s+x_1^sx_2^r) \, .
\end{equation}
Consider first the case where $r>s$, so that $f_{(r,s)}=1$. Since
the action of $1-\kappa_{12}$ vanishes in absence of $\t$ terms,
the action of $\mc{C}_{12}$ is simply
\begin{eqnarray}
\mc{C}\, m_{(r,s)} = \mc{C}_{12}m_{(r,s)} &=&
r\sum_{\ell=0}^{r-s-2}x_1^{r-2-\ell}x_2^{s+\ell}-
s\sum_{\ell=0}^{r-s}x_1^{r-1-\ell}x_2^{s-1+\ell}\cr
       &=& r\sum_{\ell=0}^{\lfloor(r-s-2)/2\rfloor}m_{(r-2-\ell,s+\ell)}-
s\sum_{\ell=0}^{\lfloor(r-s)/2\rfloor} m_{(r-1-\ell,s-1+\ell)}\cr
&=& -sm_{r-1,s-1}+ (r-s) \sum_{\ell=1}^{\lfloor(r-s)/2\rfloor}
m_{(r-1-\ell,s-1+\ell)}
\end{eqnarray}
For $r=s$, we find exactly the same expression. Note that in
reading the result, we must set $m_{(r',s')}=0$ if one label is
negative.

\subsection{Summarizing the action of \mc{C}; the
$V_{ij}^{(\ell)}$ ladder operator}

Before giving the general action of $\mc{C}$, let us introduce the
following nomenclature for a pair $(i,j)$ of indices associated to
a superpartition $\La$:
\begin{equation}
\begin{array}{lll}
\mbox{type I}&: & \mbox{if} \quad i,j \in \{1,\dots,m \} \,,\cr
\mbox{type II}&: & \mbox{if} \quad i \in \{1,\dots,m\} \, , j \in
\{m+1,\dots,N \}\,,\cr
     \mbox{type III}&:& \mbox{if}\quad i,j \in
\{m+1,\dots,N \}\, ,
\end{array}
\label{type}
\end{equation}
where $m=\overline{\underline{\La}}$ is the fermionic sector.

     If we denote generically by $m_{[r,s]}$
the $N=2$ monomial appropriate to each of the three cases, we have
obtained:
\begin{equation}\label{actionC2}
\mc{C}\, m_{[r,s]} = \sum_{\ell=0}^{\Delta_{r,s}}
[(1-\delta_{\ell,0} ){\rm max}(r,s)-{\rm min}(r,s)-\eta \ell] \,
m_{[r-1-\epsilon\ell,s-1+\epsilon\ell]}
\end{equation}
where $\epsilon=$ sgn$(r-s)$,
\begin{equation} \label{deldefi}
\Delta_{r,s} =
\begin{cases}
\lfloor \frac{r-s-1}{2}\rfloor & {\text{for type I}} \\
\max\bigl(\, |r-s| -1,0 \bigr)  & {\text{for type II}}  \\
\lfloor \frac{r-s}{2}\rfloor &  {\text{for type III}}
\end{cases} \, ,
\end{equation}
and
\begin{equation}\label{etade} \eta=\left\{
\begin{array}{ll}
2 & \quad {\rm for}\; (i,j) \, \, {\rm{of~type~I}}\,  \cr 1  &\quad
{\rm for}\;  (i,j) \,\, {\rm{of~type~II}}\,  \cr 0 & \quad {\rm
for}\; (i,j) \,\, {\rm{of~type~III}} \, \cr
\end{array}
\right. \, .
\end{equation}

The action of the operator $\mc{C}$ on supermonomials can be
decomposed into three factors: a numerical prefactor (computed in
the $N=2$ sector), a symmetry factor (see below) and the action of
a second {\it ladder operator} $V_{ij}^{(\ell)}$ acting on
superpartitions.  Its action
      is  analogous to that of the $S_{ij}^{(\ell)}$
operator, presented  in Appendix A,  pertaining to the description
of the Jack superpolynomials. For $i<j$, it is defined only for $0\leq
\ell\leq \Delta_{\La_i\La_j}$, in which case
\begin{equation}
V_{ij}^{(\ell)} \La = V_{ij}^{(\ell)} (\La_1,\cdots, \La_i,\cdots,
\La_j,\cdots)= (\La_1,\cdots, \La_i-1-\epsilon\ell,\cdots,
\La_j-1+\epsilon\ell,\cdots) \, ,
\end{equation}
where $\epsilon$ is now given by
\begin{equation}\label{dudu}
\epsilon= {\rm sgn}(\La_i-\La_j) \, .
\end{equation}
Note that here, in
contradistinction with the action of the operator
$S_{ij}^{(\ell)}$, the number $\ell$ can be zero. The possible
sign resulting from the reordering of
$\overline{V_{ij}^{(\ell)}\Lambda}$ will be written
${\rm{sgn}}(\sigma^a_{V_{ij}^{(\ell)}\Lambda})$.

We now take into account the symmetry factors.  They have been
evaluated in Section 4.2 of Ref. \cite{DLM2} for the case of Jack
superpolynomials. However, such symmetry factors are universal:
they can be used for any ladder operator $\mc{L}_{ij}^\ell$ whose
action on a superpartition $\La$ is given by $(\La_1,\cdots,
\La_i\pm a-\epsilon\ell,\cdots, \La_j\pm a+\epsilon\ell,\cdots)$
where $a$ is a positive integer. Consequently, combining the $N=2$
contribution (\ref{actionC2}) and the known symmetry factors, we
get
\begin{equation}\label{actionCgen}
\mc{C}\, m_{\La}=\sum_{\Om}e_{\La \Om}\,m_\Om\, ,\qquad e_{\Lambda
\Omega} = \sum_{(i,j) \text{~distinct}; \overline{V_{ij}^{(\ell)}
\Lambda}=\Omega} \tilde{e}_{\Lambda,V_{ij}^{(\ell)} \Lambda } \, ,
\end{equation}
where
\begin{equation}\label{deldef}
      \tilde{e}_{\Lambda, V_{ij}^{(\ell)} \Lambda}=
[{\rm max}(\La_i,\La_j)(1-\delta_{\ell,0})-{\rm
min}(\La_i,\La_j)-\eta \ell]\, n(\La_i-1-\epsilon \ell,
\La_j-1+\epsilon \ell).
\end{equation}
The parameter $\eta$ is defined in \eqref{etade} and
     $ n(a,b) $  is given by:
\begin{equation}\label{symfa}
n(a,b)=\left\{
\begin{array}{ll}
1 & {\rm for}\; i,j \, \, {\rm{of~type~I}}\,  \cr n_{\Omega^s}(b)
&{\rm for}\;  i,j \,\, {\rm{of~type~II}}\,  \cr
n_{\Omega^s}(a)n_{\Omega^s}(b) & {\rm for}\; i,j \,\,
{\rm{of~type~III~and}} \, \, a \neq b \, \cr \frac{1}{2}
n_{\Omega^s}(a)\left(n_{\Omega^s}(a)-1\right) & {\rm for}\; i,j
\,\, {\rm{of~type~III~and}} \, \, a= b \, \cr
\end{array}
\right.
        \end{equation}
where  as before, $n_{\Om^s}(i)$ denotes the number of $i$'s in
$\Om^s$, the symmetric part of $\Omega=\overline{ V_{ij}^{(\ell)}\Lambda}$.

In \eqref{actionCgen}, we say that $(i,j)$ and $(i',j')$ are
distinct if $(\La_i, \La_{j}) \neq (\La_{i'}, \La_{j'})$ or if
$\La_i$ and $\La_{i'}$ belong to different constituent partitions
$\La^a$ or $\La^s$. The sum over the pairs $(i,j)$ is required by
the fact that a given term can be obtained in different ways, some
of them associated to different symmetry factors. For example, let
the initial and the final superpartitions be $\La=(0;2,1)$ and
$\Om=(0;1)$ respectively. Three distinct pairs $(i,j)$ contribute in
\eqref{actionCgen}: $(1,2)$ for $\ell=1$, $(2,3)$ for $\ell=0$ and
$(2,4)$ for $\ell=1$.  Formulas \eqref{deldef} and \eqref{symfa}
yield directly: \beq \tilde{e}_{\La, V_{12}^{(1)}\La}=(N-2)\,
,\qquad \tilde{e}_{\La, V_{23}^{(0)}\La}=-(N-2)\, ,\qquad
\tilde{e}_{\La, V_{24}^{(1)}\La}=(N-2)(N-3)\, ,\eeq so that
\beq\mc{C}\, m_{(0;2,1)}=(N-2)(N-3)\, m_{(0;1)}\, .\eeq
In addition, this is also an example for
which $\Om$ is linked to
$\La$ by the `one-body ladder operator' $U_2$, -- cf. the example at the
end of Section 3.1.

We stress
that this pattern for the action of $\mc{C}$, which describes the
interacting part of the Hamiltonian, would have been impossible to
unravel without going through the $N=2$ analysis.

\subsection{Complete action of the Hamiltonian; the $v$-ordering}

Combining the above expressions for the action of $\mc{A}$,
$\mc{B}$ and $\mc{C}$, we get the following compact expression for
the action of the Hamiltonian:
\begin{equation}\label{hut}
\mc{H}\, m_{\La}=2\om \, (
|\La|+\overline{\underline{\La}})\,m_\La-\sum_{\Om;\Om=\overline{U_i\La}}
d_{\La\Om}\, m_\Om
-2\beta\,\sum_{\Om;\Om=\overline{V_{ij}^{(\ell)}\La}}e_{\La\Om}\,
m_\Om\, ,
\end{equation}
where the coefficients $d_{\La,\Om}$ and $e_{\La,\Om}$ are given
by (\ref{pLaOm}) and \eqref{actionCgen}-\eqref{deldef}
respectively.

The action of the srCMS Hamiltonian on the supermonomial basis
induces a particular ordering on superpartitions that enters in
the triangular decomposition of the generalized Hermite
polynomials. Let us denote this ordering as $\leq_v$. We have seen
that the action  of $\mc{H}^\om$ on $\, m_{\La}$ is triangular and
that the `lower-order' superpartitions are simply those such that
$\Om=\overline{U_i\La}$ or $\Om= \overline{V_{ij}^{(\ell)}
\Lambda}$ for some $i,j$. In the triangular decomposition of
$\mc{H}^\om$ on $m_{\La}$, only terms obtained by one application
of either $U_i$ or $V_{ij}^{(\ell)} $ can appear. It is thus
natural to define the $v$-ordering $\leq_v$ in terms of multiple
applications of these operators: \beq
\Om\leq_v\La\qquad\mbox{iff}\qquad
\Om=\overline{V_{I_1}\ldots\overline{V_{I_n}\La}}\eeq
     for a given
sequence $V_{I_1},\ldots,V_{I_n}$, where $V_{I_k}$ stands for the
ladder operator $U_{i_k}$ or $V_{i_k j_k}^{(\ell_k)}$ .

Again, the combined action of ladder operators $U_i$ and
$V_{ij}^{(\ell)}$ on a superpartition $\Lambda$ decreases his
weight $|\Lambda|$.  This means that the $v$-ordering furnishes
another  refinement of the $u$-ordering: \beq\label{lemmavorder}
\mbox{if}\qquad \Om\leq_v\La\qquad\mbox{then}\qquad\Om\leq_u\La\,
.\eeq

     To summarize, the action of $\mc{H}$ is triangular
in the supermonomial basis $\{m_\La\}_\La$ : \beq\label{developH}
\mc{H}\,
m_\La=2\om \, (|\La|+\underline{\overline{\La}})\,m_\La+\sum_{\Om <_v
\La}c_{\La\Om}(\beta,\om,N) \, m_\Om, \eeq with
\begin{equation}
\sum_{\Omega <_v\Lambda}c_{\Lambda\Omega}(\beta,\om,N)
\,m_{\Omega}= -\sum_{\Om; \atop\Om=\overline{U_i\La}}
d_{\La\Om}\,m_\Om \quad-\quad 2\beta \sum_{\Om}
\sum_{ (i,j)
\text{~distinct} ;
    \atop\Omega= \overline{V_{ij}^{(\ell)} \Lambda}}
      \tilde{e}_{\Lambda , V_{ij}^{(\ell)} \Lambda}\,  m_\Om \, .
\end{equation}
Note
that two superpartitions such that $\Om<_v\La$
have distinct eigenvalues, that is, $e_{\Om} \neq e_{\La}$
since $|\Om|<|\La|$ in that case. This property is essential when writing
the eigenfunctions as determinants.

\section{Determinantal formulas}

The preceding section contains three essential properties
concerning the action of the Hamiltonian $\mc{H}$ on a generic
supermonomial $m_\La$:
\begin{enumerate}\item it is finite;
\item it is triangular with respect to the $v$-ordering ;\item the
coefficient $c_{\La\Om}$ of the monomial $m_\Om$ appearing in the
development \eqref{developH} is known, that is, it can be computed
via (\ref{pLaOm}) and \eqref{actionCgen}-\eqref{deldef}.
\end{enumerate}
Thus, it is possible to construct an eigenfunction of $\mc{H}$
that has the following triangular form: \beq
m_\La+\sum_{\Om<_v\La}w_{\La\Om}(\beta,\om,N)m_\Om. \eeq But, by
virtue of Eqs \eqref{defmomega} and \eqref{lemmavorder}, this
function must be the very superpolynomial $m_\La^\om$. Moreover,
the triangularity together with the property $\Om<_v\La\Rightarrow
e_{\Om} \neq e_{\La}$ allow us to write the eigenfunction $m_\La^\om$ as a
determinant.  More precisely, using arguments similar to those in
\cite{LLM}, we get the following two results.
\begin{theorem}\label{detmomega}
Let $\Lambda^{(1)}\prec_v\Lambda^{(2)}\prec_v\ldots\prec_v
\Lambda^{(n)}=\Lambda$ where $\prec_v$ is a total ordering
compatible with $<_v$. Then the Hamiltonian $\mc{H}$ has a
triangular eigenfunction
       $m_{\Lambda}^\om$ of eigenvalue $e_\Lambda=2\omega \,
(|\La|+\overline{\underline{\La}})$ that is given by the
following determinant:
\begin{equation} \label{detomegaeq}
m_{\Lambda}^\om = {W_{\Lambda}} \left|
\begin{array}{cccccc} m_{\Lambda^{(1)}} &
m_{\Lambda^{(2)}} &\cdots & \cdots & m_{\Lambda^{(n-1)}} &
m_{\Lambda^{(n)}} \cr e_\Lambda^{(1)}-e_\Lambda^{(n)} &
c_{\Lambda^{(2)}\Lambda^{(1)}}&\cdots & \cdots &
c_{\Lambda^{(n-1)}\Lambda^{(1)}} & c_{\Lambda^{(n)}\Lambda^{(1)}}
\cr
                                         0&
e_\Lambda^{(2)}-e_\Lambda^{(n)}& \cdots& \cdots &
c_{\Lambda^{(n-1)}\Lambda^{(2)}}
        & c_{\Lambda^{(n)}\Lambda^{(2)}} \cr
                       \vdots &0 &\ddots &&\vdots& \vdots \cr
                       \vdots &\vdots &\ddots &\ddots& & \vdots \cr
0 & 0 &\cdots & 0 & e_\Lambda^{(n-1)}-e_\Lambda^{(n)}  &
c_{\Lambda^{(n)}\Lambda^{(n-1)}}
\end{array} \right| \, ,
\end{equation}
where the `weight' constant of proportionality is
\begin{equation}
W_{\Lambda} = (-1)^{n-1} \prod_{i=1}^{n-1}
\frac{1}{e_\Lambda^{(i)}-e_\Lambda^{(n)}} \, .
\end{equation}
\end{theorem}

\begin{corollary}
Let $\Lambda^{(1)}\prec_v\Lambda^{(2)}\prec_v\ldots\prec_v
\Lambda^{(n)}=\Lambda$ where $\prec_v$ is a total ordering
compatible with $<_v$. Then \beq
m^\om_\La\,=\,\sum_{k=1}^n
w_{\La\La^{(k)}}(\beta,\om,N)\, m_{\La^{(k)}},\eeq where
$w_{\La\La^{(n)}}=w_{\La\La}=1$, and where the coefficients
$w_{\La\La^{(k)}}$ for $1< k\leq n$ satisfy the following
recursion formula: \beq
w_{\La\La^{(k-1)}}=\frac{1}{e_\La-e_\La^{(k-1)}}\sum_{\ell=k}^{n}
w_{\La\La^{(\ell)}}c_{\La^{(\ell)}\La^{(k-1)}}.\eeq
\end{corollary}

However, the eigenfunctions $m_\La^\om$ are not orthogonal with
respect to the scalar product $\langle\, ,\,\rangle_{\beta,\om}$.
Furthermore, in the zero-fermion sector, the dominant term in
$m^\om_\La=m_\La+\sum_{\Om<_v\La}w_{\La\Om}\,m_\Om$ is not the
Jack polynomial $J_\La$. Consequently, the function $m^\om_\La$
cannot be the sought for generalized Hermite polynomial
$J^\om_\La$. The appropriate way to construct the generalized
Hermite polynomials in superspace is however obvious at this
point: linearly combine the eigenfunctions $m^\om_\Gamma$ sharing
the same weight $|\La|$ in such a way that the Jack
superpolynomial $J_\La$ is the term of homogeneous degree $|\La|$
of this linear combination.

\begin{theorem}\label{theofinal}Let $\leq$ be the dominance ordering
on superpartitions (cf. Eq. \eqref{dominanceorder}),
     and let  $t_{\La\Gamma}$ be  the coefficient of $m_\Gamma$ in the
development of the Jack polynomial $J_\La$ given in Eq.
(\ref{defjack1}). Then, the generalized Hermite polynomial in
superspace $J_\La^\om$ reads \beq\label{tridet}
J^\om_\La=m_\La^\om+\sum_{\Gamma<\La}t_{\La\Gamma}(\beta)m^\om_\Gamma.\eeq
\end{theorem}
\n{\it Proof.} First, it is obvious that the expression
(\ref{tridet}) defines an eigenfunction of the Hamiltonian
$\mc{H}$ with eigenvalue $2\om \, (|\La|+ \overline{\underline{\La}})$.
Second, let us recall that the generalized Hermite superpolynomials
$J_\La^\om$ are the unique eigenfunctions of $\mc{H}$ in superspace such
that (cf. Eq.
\eqref{defhermite1}) $J^\om_\La =J_\La+\sum_{\Om<_u\La}w_{\La
\Om}J_\Om$.  But Eqs (\ref{tridet}),  \eqref{defjack1} and
\eqref{uorder} also imply that \beq
m_\La^\om+\sum_{\Gamma<\La}t_{\La\Gamma}(\beta)m^\om_\Gamma
=J_\La+\sum_{\Gamma<_u\La}\tilde{t}_{\La\Gamma}(\beta,\om,N)J_\Om,\eeq
for some coefficient $\tilde{t}_{\La\Gamma}$.  Thus, from the
uniqueness property,
$m_\La^\om+\sum_{\Gamma<\La}t_{\La,\Gamma}m^\om_\Gamma$ must be
the generalized Hermite superpolynomial $J^\om_\La$. \hfill
$\square$

The equation (\ref{tridet}) can be interpreted as a
`tri-determinantal' formula  of the generalized Hermite
polynomials in terms of the supermonomials $m_\Om$.  Why? Because
the coefficients $t_{\La\Om}$ come from determinants
(\ref{detsjack1}) and (\ref{detsjack2})  while the coefficients
$w_{\La\Om}$ are obtained from determinant (\ref{detomegaeq}).
    From the computational point of view, this means that, when the
monomial decomposition of a Jack superpolynomial $J_\La$ is known
(using a bi-determinantal formula), we get the exact expression of
$J_\La^\om$ in the monomial basis by developing the
$m_\Gamma^\om$'s in (\ref{tridet}) via the  determinantal formula
(\ref{detomegaeq}).\\

Before concluding this article, it might be of interest  to
present an example.  Let us compute $J^\om_{(0;2,1)}$ in terms of
the Jack superpolynomials using the determinantal formulas.
Tables 1 and 2 in Ref. \cite{DLM3} (or the determinantal formulas
in Appendix A) give the Jack superpolynomial \beq
J_{(0;2,1)}=m_{(0;2,1)}+\frac{2\beta}{1+2\beta}
m_{(1;1^2)}+\frac{6\beta}{1+2\beta}m_{(0;1^3)}.\eeq
Theorem \ref{theofinal} tells us that the generalized Hermite
superpolynomial associated to $\La=(0;2,1)$ is simply
\beq\label{exHermite}
J_{(0;2,1)}^\om=m_{(0;2,1)}^\om+
\frac{2\beta}{1+2\beta}m_{(1;1^2)}^\om+\frac{6\beta}{1+2\beta}m_{(0;1^3)}^\om.\eeq
We next compute the eigenfunctions $m_{(0;2,1)}^\om$,
$m_{(1;1^2)}^\om$ and $m_{(0;1^3)}^\om$ in the supermonomial
basis using Theorem \ref{detmomega}.  Direct
calculations give
\beq\begin{split}
     &m_{(0;2,1)}^\om&=\quad
&m_{(0;2,1)}-\frac{(N-2)[1+\beta(N-3)]}{2\om}\,m_{(0;1)}\,,\cr
&m_{(1;1^2)}^\om&=\quad &m_{(1;1^2)}+\frac{\beta(N-2)}{2\om}\,
m_{(0;1)}+\frac{\beta(N-1)(N-2)}{4\om}m_{(1;0)}\, ,\cr
&m_{(0;1^3)}^\om&=\quad &
m_{(0;1^3)}+\frac{\beta(N-2)(N-3)}{4\om}\, \ds m_{(0;1)}\, ,\cr
     &m_{(1;0)}^\om&=\quad &m_{(1;0)}\,=\,
J_{(1;0)}-\frac{\beta}{1+\beta}J_{(0;1)}\, ,\cr
     &m_{(0;1)}^\om
&=\quad & m_{(0;1)}\, =\, J_{(0;1)}\, .
     \end{split}\eeq
Substituting these expansions into \eqref{exHermite} leads to the
decomposition of $J_{(0;2,1)}^\om$ in the undeformed supermonomial basis.
Although we do not have derived in general the explicit form of the
decomposition in terms of Jack superpolynomials,
in this special case it can be easily checked to read as:
   \beq  \label{exHerJ}
J_{(0;2,1)}^\om=
   J_{(0;2,1)}+\frac{\beta^2(N-1)(N-2)}{2\om(1+2\beta)}J_{(1;0)}
   -\frac{(N-2)(1+N\beta)}{2\om(1+\beta)(1+2\beta)}J_{(0;1)}\,.\eeq

\section{Conclusion}

We have expressed the generalized Hermite superpolynomials,
the eigenfunctions of the srCMS model, in determinantal form. This
amounts to writing them  in terms of the deformed
supermonomial basis as
\begin{equation}
J^\om_\La=m_\La^\om+\sum_{\Gamma<\La}t_{\La\Gamma}(\beta)\,m^\om_\Gamma
\end{equation}
with known coefficients. Notice that in this
decomposition,  all the dependence on $N$ comes from the
$m_\La^\om$'s. This nevertheless implies that, in
contradistinction with the Jack superpolynomials, the dependence
on $N$ is inherent in the generalized Hermite superpolynomials. In
other words, the
$J^\om_\La$'s are not stable with respect to the number of variables.
\footnote{See also
the example \eqref{exHerJ}) which express a special $J^\om_\La$ in
the $J_\La$ basis; in that case, the $N$-dependence is completely captured
by the coefficients.}

The somewhat complicated form of the expressions resulting from
this construction should not cast a shadow over the fact that all
eigenfunctions of the srCMS model are thereby obtained in an
explicit way.

\begin{appendix}

\section{A general scheme for the explicit construction of the
eigenfunctions}

A large class of generalized orthogonal symmetric functions  can
be characterized by an eigenvalue problem of the
Calogero-Moser-Sutherland (CMS) type. More precisely, these
multi-variable polynomials are eigenfunctions of a gauged CMS
Hamiltonian $\cal H$,
      where the gauge transformation
refers to the removal of the ground-state wave function $\psi_0$
from the genuine CMS  Hamiltonian $H$ via the conjugation
${\cal H}= \psi_0 H
\psi_0^{-1}$. A crucial property of $\cal H$ is that it can be
broken up as a sum of  one- or two-body interaction terms, i.e., as
\beq {\cal H}= \sum_{i,j=1}^N {\cal H}_{ij}= \sum_{i,j=1}^N [{\cal
H}_{ij}^{(1)} + {\cal H}_{ij}^{(2)}+\cdots]\, , \eeq
with   $N$ the number of interacting particles. Here ${\cal
H}_{ij}^{(p)}$ could be a diagonal term (e.g., a kinetic energy
term)  or more generally a single-body term, in which case it would be
proportional to
$\delta_{ij}$, or an interaction term, in which case ${\cal
H}_{ij}^{(p)}\propto (1-\delta_{ij})$.

There is a systematic scheme for constructing explicitly the
eigenfunctions of ${\cal H}$ that also  provides closed-form
expressions.  By an explicit construction, we refer to the
complete specification of the expansion coefficients of the sought
for eigenfunctions
      in a prescribed basis, typically the basis of
monomial symmetric functions $m_\la$, where $\la$ is  a partition.
More precisely, we look for  eigenfunctions $P_\la$
      of the form
\beq \label{deco}
      P_\la= m_\la+ \sum_{\mu < \la} \alpha_{\la,\mu} m_\mu \, ,
\eeq that is, triangular with respect to a certain ordering
on partitions and normalized such that
$\alpha_{\la,\la}=1$.
In short, the explicit construction of $P_\la$ refers to the
determination of the coefficients $\alpha_{\la,\mu}$.

This scheme relies heavily on the intrinsic two-body nature of the
CMS interaction. The first step amounts to treat the $N=2$ problem
and evaluate explicitly the action of ${\cal H}_{12}$ on a generic
(two-part) monomial function $m_\la= m_{(r,s)}$ $(r=\la_1,\,
s=\la_2)$: \beq {\cal H}_{12} m_{(r,s)}=\sum_p{\cal H}_{12}^{(p)}
m_{(r,s)}= \epsilon_{(r,s)}\, m_{(r,s)}+
\sum_{(r',s')\not=(r,s)}\sum_p c_{(r,s),(r',s')}^{(p)} \,
m_{(r',s')} \eeq
In
the CMS case, the
action actually turns out to be triangular:  the
ordering $<$ on partitions with two entries being fixed
such that $(r',s')<(r,s)$ when $(r',s')$
appears in the sum.

Actually, to each element ${\cal H}_{12}^{(p)}$ in the
decomposition of ${\cal H}_{12}$ corresponds a ladder
(lowering) operator ${\cal L}_{12}^{(p)}$ acting on partitions.
To be more precise, to any $m_{(r',s')}$ appearing in ${\cal
H}_{12} m_{(r,s)}$ there corresponds a ladder operator that acts
on the partition $(r,s)$ to produce $(r',s')$. The ordering
governing the triangular decomposition can thus be defined
precisely as follows: $(r',s')<(r,s)$ if there is a ladder
operator ${\cal L}_{12}^{(p)}$ such that $(r',s')= {\cal
L}_{12}^{(p)}(r,s)$.

This first step which consists in solving  the $N=2$ case, even though
clearly model-dependent, is usually rather straightforward.
The second step now amounts to extending this result from $N=2$ to a
general $N$, that is, to obtaining explicitly
    \beq {\cal H} m_{\la}=
\epsilon_{\la}\, m_{\la}+ \sum_{\mu<\la} c_{\la,\mu}\, m_{\mu}\, .
\eeq The computation of the diagonal coefficient $\epsilon_{\la}$
is normally quite simple. The main technical difficulty lies in the
determination of the
non-diagonal coefficients $c_{\la,\mu}$. The key point is that all
the monomials $m_\mu$  that appear in this expansion  can be
traced back to an underlying two-body interaction. In other words,
a partition $\mu$ such that $c_{\la,\mu}\not=0$
      differs from $\la$ in at most two of its entries, say the $i$-th and
the $j$-th ones.  This implies that $c_{\la,\mu}$ is determined by
the corresponding two-body coefficient
$c_{(\la_i,\la_j),(\mu_i,\mu_j)}^{(p)}$
      up
to a symmetry factor $a_\mu^{(p)}$: \beq
      c_{\la,\mu}= \sum_p a_\mu^{(p)}
      c_{(\la_i,\la_j),(\mu_i,\mu_j)}^{(p)}\, .
\eeq This symmetry factor $a_\mu^{(p)}$ is characteristic of the
generic form of ${\cal H}_{ij}^{(p)}$, that is, whether it is
a single- or two-body term. It takes care of the different
ways we can relate $\la$ to $\mu$; it is thus a purely combinatorial
factor, solely determined by the multiplicity of the part $\mu_i$
in $\mu$ if it is related to a diagonal term, or the multiplicity
of both $\mu_i$ and $\mu_j$ for a genuine interaction term.  In
that sense, the symmetry factors are
      universal, i.e., independent of the precise form of the
Hamiltonian ${\cal H}$.
\footnote{ \label{foot1}  To be more precise, let us stress that in
general, we may also have
to sum over pairs $(\lambda_i,\lambda_j)$ that are
combinatorially distinct in the sense of belonging to separate
families, each of which giving rise to
different symmetry factors (see the paragraph following
\eqref{symfa} for an explicit example).  However, the families are
only determined by the structure of the partitions involved (the
superpartitions in the case treated in this paper), and it thus remains
true that the combinatorial factor is independent of the structure of the
Hamiltonian. Note that the case we consider in this paper captures the
full complexity of the generic situation described above in that  a
  given monomial term $m_\Om$ occurring in the decomposition of
  ${\cal H} m_\La$ can be related to two different pieces of the
two-body Hamiltonian, or to combinatorially distinct pairs
  $(\lambda_i,\lambda_j)$  associated to a given two-body term. }

After the completion of these first two steps, the action of
${\cal H} \, m_{\la}$ is known. The third step amounts to a direct
construction of the eigenfunctions $P_\la$ in the form of a
determinant involving the following entries: $m_\la, \epsilon_\la$
and $ c_{\la,\mu}$. By expanding the determinant, we obtain the
decomposition (\ref{deco}) with all $\alpha_{\la,\mu}$ determined.
The sum runs now over all $\mu<\la$, where the ordering governing
this triangular decomposition is  precisely the ordering that
underlies the decomposition of ${\cal H} m_{\la}$ (in the sense
that we say $\mu< \lambda$ if $m_{\mu}$ appears in ${\cal H}^n
m_{\la}$, for a certain value of $n$).

The general scheme is thus quite simple in principle and totally
constructive. Explicit eigenfunctions can be worked out in this
way without requiring the previous knowledge of anything more than
the monomial basis. Examples of orthogonal polynomial that can be
computed in this way include the Jack polynomials and the
generalized Hermite, Jacobi and Laguerre polynomials.
For Jack polynomials, eigenfunctions of the trigonometric CMS
(tCMS) model,
the above procedure is particularly simple
because there is a single two-body term in the Hamiltonian that
contributes to the coefficient $c_{\la,\mu}$ and also a
single overall symmetry factor.

It is difficult to trace back the precise origin of this scheme in
the literature. To our knowledge, its `history' goes as follows.
For Jack polynomials $J_\la$, it appears implicitly in Macdonald's
book \cite{Ma}. In particular,  the action of ${\cal H} m_{\la}$
is given  in example 3(c) of Sect. VI.4 p. 327. In  this
expression we can recognize the two-body decomposition but there
is no emphasis on the symmetry factor, which is hidden in the form of
a summation.
Therein, the
following example 3(d)
      displays a recursion formula for the coefficients
$\alpha_{\la,\mu}$ that can also be deduced from  the
determinantal expression of $J_\la$.  The expression of ${\cal H}
m_{\la}$ for the tCMS model, using the idea of first computing the
two-body case and then lifting the result by a symmetry factor is
due to Sogo \cite{So} (but given without proof). The completion of
the program by displaying the explicit determinantal expression
for the eigenfunctions (in this case, the Jack polynomials)  has
first been done in \cite{LLM}.  Furthermore, the extension to all
root systems was considered in \cite{FLJ}.

The first explicit formulation of the above complete scheme
appears to be that of \cite{DLM2} (see also Sect. 3.6-3.8 of
\cite{DLM4}), where it was applied to the construction of the Jack
superpolynomials, defined to be eigenfunctions of the
supersymmetric version of the tCSM model \cite{DLM1}.

In the absence of supersymmetry, the above construction leads
directly to orthogonal polynomials. Indeed, eigenfunctions of
${\cal H}$ turn out to be also eigenfunctions of a whole tower of
commuting charges $\{{\cal H}_n\}$, for $n=1,2,\dots,N$, (where
${\cal H}_2={\cal H}$) since these charges also act triangularly
on the monomial basis with respect to the same ordering. However,
in the supercase,  it is not guaranteed that the superpolynomials,
denoted ${\cal P}_\La$, constructed along this scheme will be
orthogonal. And actually, this is usually not the case. But the
way one should linearly combine the ${\cal P}_\La$'s to form
orthogonal superpolynomials is suggested by the integrability
structure of the underlying physical problem. It is indeed rooted
in the fact that the supersymmetric extension of any CMS model has
twice as many commuting conserved charges than its corresponding
non-supersymmetric version.  In other words, there is an
additional tower of $N$ conserved charges $\{{\cal I}_n\}$ (which
vanish when the anticommuting variables are set equal to zero).
One thus needs to iterate the procedure one step further: take a
representative among this new set of conserved charge (in its
gauged form) to play the role of $\cal H$ and replace $m_\La$ (the
supermonomial) by the eigenfunction ${\cal P}_\La$ that was built
in the first step using $\cal H$. The resulting eigenfunctions are
then orthogonal.  This is precisely the way the orthogonal Jack
superpolynomials $J_\La$ have been obtained in \cite{DLM3}.

\section{Determinantal formulas for Jack polynomials in superspace}

The explicit formulas for the Jack polynomials $J_\La$ in
superspace are written in terms of two bases of $P^{S_N}$: the
non-orthogonal set of eigenfunctions $\{ \mc{J}_\La\}_\La$ and the
monomial set $\{m_\La\}_\La$.  These formulas make use of two
special orderings on superpartitions.

The first partial ordering in related to the action of a ladder
operator $ S_{ij}^{(\ell)} $ on superpartitions.\footnote{The
operator $ S_{ij}^{(\ell)} $ reads $ R_{ij}^{(\ell)} $ in
\cite{DLM2}.} Its action, for $ i<j $ and $ \ell \geq 1$, is
defined as follows:
\begin{equation} \label{sij}
S_{ij}^{(\ell)}(\Lambda_1,\dots,\Lambda_i,\dots,\Lambda_j,\dots)
=\left\{
\begin{array}{ll}
(\Lambda_1,\dots,\Lambda_i-\ell,\dots,\Lambda_j+\ell,\dots) &
{\rm{if }} \, \, \Lambda_i > \Lambda_j\,, \cr
(\Lambda_1,\dots,\Lambda_i+\ell,\dots,\Lambda_j-\ell,\dots) &
{\rm{if }} \, \, \Lambda_j > \Lambda_i\,.
\end{array}\right.
\end{equation}
Note that the result is such as given only if :
\begin{equation}
\begin{array}{rl}
{\rm{I}}: & \quad i,j \in \{1,\dots,m \}\,  \quad {\rm{and}} \quad
\lfloor \frac{\Lambda_i-\Lambda_j-1}{2}\rfloor \geq \ell \,,\cr
{\rm{II}}: & \quad i \in \{1,\dots,m\} \, , j \in \{m+1,\dots,N
\}\, \quad {\rm{and}} \quad |\Lambda_i-\Lambda_j|-1 \geq
\ell\,,\cr {\rm{III}}: & \quad i,j \in \{m+1,\dots,N \}\,  \quad
{\rm{and}} \quad \lfloor \frac{\Lambda_i-\Lambda_j}{2}\rfloor \geq
\ell \, , \cr
\end{array}
\end{equation}
and otherwise, $ S_{ij}^{(\ell)}\, \Lambda=\emptyset $. The {\it
$s$-ordering} $\leq_s$ is defined by: \beq
\Omega\leq_s\Lambda\qquad \mbox{iff}\qquad\Omega= \overline{S_{i_k
j_k}^{(\ell_k)} \dots \overline{S_{i_1j_1}^{(\ell_1)} \Lambda}}
\eeq for a given sequence of operators    $ S_{i_1
j_1}^{(\ell_1)},\dots,S_{i_k j_k}^{(\ell_k)} $.

Obviously, if $\Omega\leq_s\Lambda$ then $\Omega\leq\Lambda$,
i.e., the $s$-ordering refines the dominance ordering.  The same
property  holds for the second ordering on superpartitions.  It is
defined using the exchange operator $T_{ij}$ whose action on
$\Lambda=(\ldots,\Lambda_i,\ldots,\Lambda_j,\ldots)$ is not
influenced by the semi-colon and is given by : \beq
T_{ij}\Lambda=\begin{cases}
(\ldots,\Lambda_j,\ldots,\Lambda_i,\ldots)&\mbox{if}\quad\Lambda_i>\Lambda_j\,
,\\
(\ldots,\Lambda_i,\ldots,\Lambda_j,\ldots)\, &\mbox{otherwise
.}\end{cases} \eeq
     The {\it $t$-ordering} $\leq_t$ is  such that:\beq
     \Omega\leq_t\Lambda\qquad\mbox{iff}\qquad\Omega= \overline{T_{i_k
j_k} \dots
\overline{T_{i_1 j_1} \Lambda}}\eeq for a given sequence of
operators $ T_{i_1 j_1},\dots,T_{i_k j_k} $.

We are now in position to give  another definition of the Jack
superpolynomials in terms of the eigenvalue problem associated
with two conserved operators of the stCMS model: the
supersymmetric Hamiltonian $\mc{H}_2$ and the operator $\mc{I}_1$
(which does not exist in the non-supersymmetric case). Note that
$\lim_{\beta\rightarrow 0}\mc{H}_2=\sum_i(x_i\partial_i)^2$ while
$\lim_{\beta\rightarrow
0}\mc{I}_1=(N-1)!\sum_i(x_i\partial_i)(\theta_i\partial_{\theta_i})$.

\begin{definition} \cite{DLM3}  The Jack polynomials $J_\Lambda$ in
superspace are the
unique functions in $P^{S_N}$ such that
     \beq
      \mc{I}_1J_\Lambda=\epsilon_\La J_\Lambda
\qquad\mbox{and}\qquad
J_\Lambda=\mc{J}_\Lambda+\sum_{\Omega<_t\Lambda}u_{\Lambda\Omega}(\beta)\mc{J}_\Omega
\eeq where the superpolynomials $\mc{J}_\Lambda$  are the unique
functions such that \beq \mc{H}_2\mc{J}_\Lambda=
\varepsilon_{\Lambda}\mc{J}_\Lambda
\qquad\mbox{and}\qquad\mc{J}_\Lambda=m_\Lambda+\sum_{\Omega<_s\Lambda}v_{\Lambda\Omega}(\beta)m_\Omega.
\eeq The eigenvalues are given by \beq
     \epsilon_\La=(N-1)!\sum_{i=1}^{m}[\lambda_i-\beta(m(m-1)+\#_\Lambda)]
\eeq and\beq
\varepsilon_\La=\sum_{i=1}^N[\Lambda_i^2+\beta(N+1-2j)\Lambda]\, ,
     \eeq
where $m$ stands for the fermionic sector
$\overline{\underline{\Lambda}}$,  while $\#_\Lambda$ stands for
the number of pairs $(i,j)$ such that $i\in\{1,\ldots,m\}$,
$j\in\{m+1,\ldots,N\}$ and $\Lambda_i<\Lambda_j$.
\end{definition}

We compute the coefficients $u_{\La\Om}$ and $v_{\La\Om}$ by means
of determinantal formulas (cf. \cite{DLM2,DLM3}). These formulas
are written in terms of coefficients $a_{\Lambda\Omega}$ and
$b_{\Lambda\Omega}$ defined by the following expressions: \beq
     \mc{H}_2
m_\Lambda=\varepsilon_{\Lambda}m_\Lambda+\sum_{\Omega<_s\Lambda}a_{\Lambda\Omega}(\beta)\,m_\Omega
\qquad\mbox{and}\qquad \mc{I}_1
\mc{J}_\Lambda=\epsilon_{\Lambda}\mc{J}_\Lambda+\sum_{\Omega<_t\Lambda}b_{\Lambda
\Omega}(\beta,N)\,\mc{J}_\Omega. \eeq The coefficient
$b_{\Lambda\Omega}$ reads: \beq
b_{\Lambda\Omega}=\begin{cases}(N-1)!\beta\,
\mbox{sgn}(\sigma^a_{T_{ij}\Lambda})\,
n_{\Omega^s}(\Lambda_i)&\mbox{if}\quad
\Omega=\overline{T_{ij}\Lambda}\quad\mbox{for some}\quad
i<j,\\0&\mbox{otherwise}\end{cases}\eeq

The coefficient  $ a_{\Lambda \Omega}(\beta)$ is more elaborated.
     It is non-zero only if $\Om$ can be obtained from $\La$ by a {\it
single} action of the ladder operator, namely if $ \Omega =
(\Om^a;\Om^s)= \overline{S_{ij}^{(\ell)} \Lambda} $ , for a given
        $ S_{ij}^{(\ell)} $ with $\ell>0$, in which case it
reads:
\begin{equation}\label{vexpli}
a_{\Lambda \Omega}(\beta) = \left\{
\begin{array}{ll}
2\beta \, {{\rm{sgn}}}(\sigma_{S_{ij}^{(\ell)} \Lambda}^a)
\bigl(\Lambda_i-\Lambda_j-\eta\ell  \bigr)
n(\Lambda_i-\ell,\Lambda_j+\ell) &  {\rm{if }}\, \, \Lambda_i >
\Lambda_j \, ,\cr 2\beta  \, {{\rm{sgn}}}(\sigma_{S_{ij}^{(\ell)}
\Lambda}^a) \bigl(\Lambda_j-\Lambda_i-\eta\ell  \bigr)
n(\Lambda_i+\ell,\Lambda_j-\ell) &  {\rm{if }} \, \,\Lambda_j >
\Lambda_i\, ,
\end{array}\right.
\end{equation}
where  ${{\rm{sgn}}}(\sigma_{S_{ij}^{(\ell)} \Lambda}^a) $
       stands for the  sign of the permutation $\sigma_{S_{ij}^{(\ell)}
\Lambda}^a$ and $\eta$ is given in \eqref{etade}.

Finally, the closed form expressions of the $\mc{J}_\La$'s and the
$J_\La$'s are contained in the two following theorems.
\begin{theorem} \cite{DLM2} Let
$\Lambda^{(1)}\prec_s\Lambda^{(2)}\prec_s\ldots\prec_s
\Lambda^{(n)}=\Lambda$ where $\prec_s$ is a total ordering
compatible with $<_s$. Then
       $\mc{J}_{\Lambda}$ is given by the
following determinant:
\begin{equation}  \label{detsjack1}
\mc{J}_{\Lambda} = {\mc{E}_{\Lambda}} \left| \begin{array}{cccccc}
m_{\Lambda^{(1)}} & m_{\Lambda^{(2)}} &\cdots & \cdots &
m_{\Lambda^{(n-1)}} & m_{\Lambda^{(n)}} \cr
\varepsilon_{\Lambda^{(1)}}-\varepsilon_{\Lambda^{(n)}} &
a_{\Lambda^{(2)}\Lambda^{(1)}}&\cdots & \cdots &
a_{\Lambda^{(n-1)}\Lambda^{(1)}} & a_{\Lambda^{(n)}\Lambda^{(1)}}
\cr                                   0&
\varepsilon_{\Lambda^{(2)}}-\varepsilon_{\Lambda^{(n)}}& \cdots&
\cdots & a_{\Lambda^{(n-1)}\Lambda^{(2)}}
        & a_{\Lambda^{(n)}\Lambda^{(2)}} \cr
                       \vdots &0 &\ddots &&\vdots& \vdots \cr
                       \vdots &\vdots &\ddots &\ddots& & \vdots \cr
0 & 0 &\cdots & 0 &
\varepsilon_{\Lambda^{(n-1)}}-\varepsilon_{\Lambda^{(n)}}  &
a_{\Lambda^{(n)}\Lambda^{(n-1)}}
\end{array} \right| \, ,
\end{equation}
where the constant of proportionality is
\begin{equation}
\mc{E}_{\Lambda} = (-1)^{n-1} \prod_{i=1}^{n-1}
\frac{1}{\varepsilon_{\Lambda^{(i)}}-\varepsilon_{\Lambda^{(n)}}}
\, ,
\end{equation}
\end{theorem}

\begin{theorem} \cite{DLM3} Let
$\Lambda^{(1)}\prec_t\Lambda^{(2)}\prec_t\ldots\prec_t
\Lambda^{(n)}=\Lambda$ where $\prec_t$ is a total ordering
compatible with $<_t$. Then
       $J_{\Lambda}$ is given by the
following determinant:
\begin{equation} \label{detsjack2}
J_{\Lambda} = {E_{\Lambda}} \left| \begin{array}{cccccc}
\mc{J}_{\Lambda^{(1)}} & \mc{J}_{\Lambda^{(2)}} &\cdots & \cdots &
\mc{J}_{\Lambda^{(n-1)}} & \mc{J}_{\Lambda^{(n)}} \cr
\epsilon_{\Lambda^{(1)}}-\epsilon_{\Lambda^{(n)}} &
b_{\Lambda^{(2)}\Lambda^{(1)}}&\cdots & \cdots &
b_{\Lambda^{(n-1)}\Lambda^{(1)}} & b_{\Lambda^{(n)}\Lambda^{(1)}}
\cr
                                         0&
\epsilon_{\Lambda^{(2)}}-\epsilon_{\Lambda^{(n)}}& \cdots& \cdots
& b_{\Lambda^{(n-1)}\Lambda^{(2)}}
        & b_{\Lambda^{(n)}\Lambda^{(2)}} \cr
                       \vdots &0 &\ddots &&\vdots& \vdots \cr
                       \vdots &\vdots &\ddots &\ddots& & \vdots \cr
0 & 0 &\cdots & 0 &
\epsilon_{\Lambda^{(n-1)}}-\epsilon_{\Lambda^{(n)}}  &
b_{\Lambda^{(n)}\Lambda^{(n-1)}}
\end{array} \right| \, ,
\end{equation}
where the constant of proportionality is
\begin{equation}
E_{\Lambda} = (-1)^{n-1} \prod_{i=1}^{n-1}
\frac{1}{\epsilon_{\Lambda^{(i)}}-\epsilon_{\Lambda^{(n)}}} \, .
\end{equation}
\end{theorem}

\end{appendix}

\medskip
\medskip
\begin{acknow}
This work was  supported by NSERC and FONDECYT (Fondo Nacional de
Desarrollo Cient\'{\i}fico y Tecnol\'ogico) grant \#1030114.
P.D. is grateful to the Fondation J.A.-Vincent for a
student fellowship.
\end{acknow}

\end{document}